\documentclass[aps,prl,10pt,twocolumn,floatfix,superscriptaddress]{revtex4-2}
\usepackage{amsfonts,amsmath,graphicx,braket,bm}
\usepackage[utf8]{inputenc}
\usepackage[colorlinks, linkcolor=blue, citecolor=blue]{hyperref}
\usepackage[table,xcdraw]{xcolor}

\begin{document}

\title{Quantum Approximate Optimization Algorithm pseudo-Boltzmann states}
\author{Pablo D\'{\i}ez-Valle}
\email{pablo.diez@csic.es}
\address{Instituto de F\'{\i}sica Fundamental IFF-CSIC, Calle Serrano 113b, Madrid 28006, Spain}

\author{Diego Porras}
\address{Instituto de F\'{\i}sica Fundamental IFF-CSIC, Calle Serrano 113b, Madrid 28006, Spain}

\author{Juan Jos\'e Garc\'{\i}a-Ripoll}
\address{Instituto de F\'{\i}sica Fundamental IFF-CSIC, Calle Serrano 113b, Madrid 28006, Spain}


\newcommand{\changedsecondround}[1]{{\textcolor{black}{#1}}}
\newcommand{\changed}[1]{{\textcolor{black}{#1}}}
\newcommand{\pablosuggestion}[1]{{\textcolor{orange}{#1}}}

\begin{abstract}
In this letter, we provide analytical and numerical evidence that the single-layer Quantum Approximate Optimization Algorithm (QAOA) on universal Ising spin models produces thermal-like states. We find that these pseudo-Boltzmann states can not be efficiently simulated on classical computers according to \changedsecondround{the general state-of-the-art condition that ensures rapid mixing for Ising models}. Moreover, we observe that the temperature depends on a hidden universal correlation between the energy of a state and the covariance of other energy levels and the Hamming distances of the state to those energies.
\end{abstract}

\maketitle


\paragraph{Introduction.--} \changed{Variational quantum algorithms~\cite{Cerezo2021} are one potential tool for near term quantum computers to show a \textit{quantum advantage} over classical methods in optimization and machine learning problems.} The Quantum Approximate Optimization Algorithm (QAOA) is a particular variational framework aimed at the solution of combinatorial optimization problems~\cite{Farhi2014}. QAOA \changed{optimizes} a trial wavefunction that alternates evolution with a simple mixing Hamiltonian and the problem Hamiltonian we wish to optimize. \changedsecondround{Despite the fact that the single-layer QAOA is amenable to analytical calculations such as the computation of expectation values ~\cite{Ozaeta2020,Farhi2022}, it has been shown that} a single-layer \changed{unoptimized QAOA circuit} \changedsecondround{already} engineers a probability distribution that is classically hard to sample~\cite{Farhi2019}. \changed{However, it is unclear whether a sampling advantage implies a practical advantage in optimization, where the variational approach may experience fundamental obstructions~\cite{Akshay2020,Bravyi2020}, and there may be better classical~\cite{Hastings2019} and quantum algorithms~\cite{amaro2022}}.

In this work we connect the sampling advantage and the optimization properties, analyzing a single-layer generalization of the QAOA ansatz. \changed{We study both the optimization power---the probability to find the ground state---and the types of states it prepares---IQP-like (\textit{Instantaneous Quantum Polynomial time}~\cite{Bremner_2016})} probability distributions that are hard to sample~\cite{Farhi2019}. \changed{We show that single-layer QAOA creates \textit{pure states} with a probabiluty distribution resembling a Boltzmann state with Gaussian perturbations. This happens for models with a correlation between eigenstate energy and spin flips, such as QUBO, Max-Cut and random Ising models in arbitrary dimensions. The effective temperature of these \textit{pseudo-Boltzmann} states depends on the optimization angles, and the type and size of the problem. For optimal angles, QAOA may reach temperatures below those efficiently sampled by Markov Chain Monte Carlo (MCMC)~\cite{eldan2021spectral}}.

\changed{This letter begins by introducing three families of random Ising models, embedded on two types of random graphs. We define} a generalization of the QAOA single-layer circuit, showing numerical evidence that \changed{it creates \textit{pseudo-Boltzmann} pure states} with an average temperature that scales favorably with the problem size, enhancing the probability to find the ground state. We give a semi-analytical interpretation \changed{of the single-layer circuit as an interferometer that macroscopically shifts probability to low or high energy states, creating pseudo-Bolztmann distributions. The efficiency of this process, measured by the effective temperature, depends on a hidden correlation in energy space.} We close with a discussion of \changedsecondround{sampling hardness}\changed{, and other questions opened by this research}

\paragraph{Optimization problems.--} We consider three families of NP-hard binary combinatorial optimization problems, QUBO~\cite{Kochenberger_2004,Kochenberger_2014}, MaxCut~\cite{Nannicini_2019,Sung_2020,Harrigan_2021} and random Ising models~\cite{Barahona1982}. \changed{QUBO and MaxCut problems have an associated classical energy function defined in binary variables $\mathbf{x}\in\{0,1\}^N,$ $E_{\textnormal{QUBO}}(\mathbf{x}) = 2\sum_{i,j=1}^N x_i Q_{ij}x_j , $ and $E_{\textnormal{MaxCut}}(\mathbf{x}) = -2\sum_{i,j=1}^N x_i Q_{ij}(1-x_j) $. Optimizing these functions is equivalent to finding the minimum-energy state of an Ising Hamiltonian for $N$ spins $\bm{\sigma^z} = (2\mathbf{x} - 1)$
\begin{equation}
  \hat{E}(\sigma^z) = \sum_{i,j=1}^N \sigma^z_i \frac{1}{2} J_{ij} \sigma^z_j + \sum_{i=1}^N h_i \sigma^z_i,
  \label{eq:energy}
\end{equation}
where $J_{i\neq j} = Q_{ij}$, $J_{ii}=0$, and $h_i = \sum_j Q_{ij}$ for QUBO and $h_i =0$ for MaxCut.}

\begin{figure}
\includegraphics[width=1\linewidth]{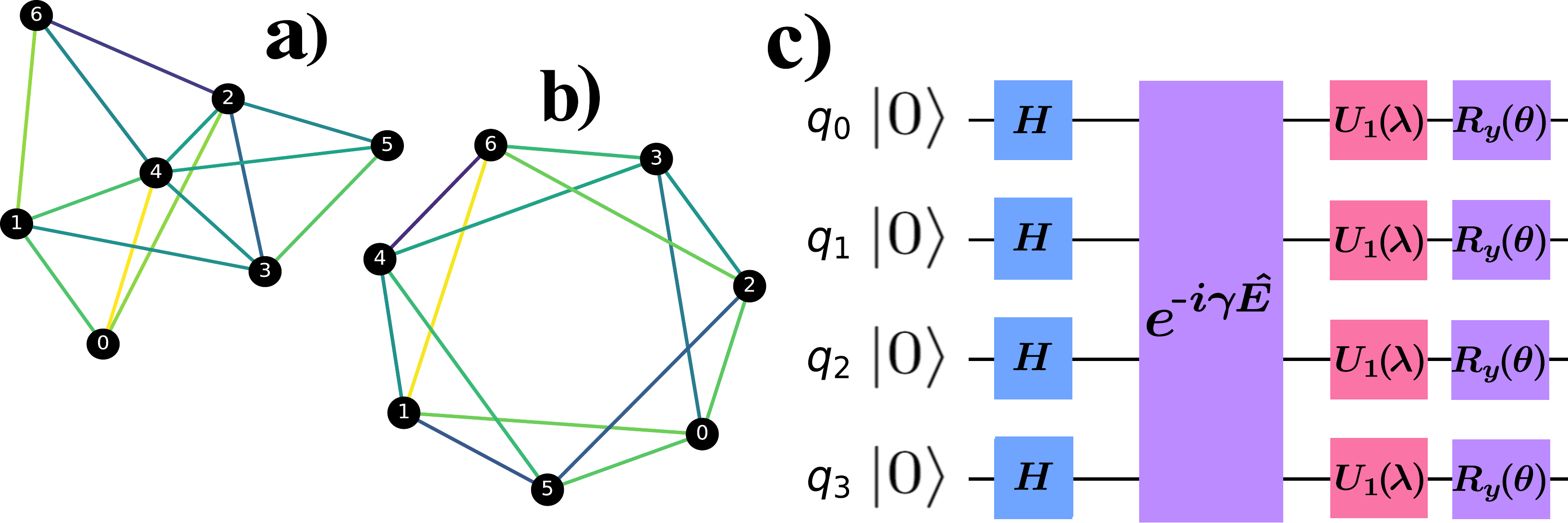}
\caption{(a) Example of a \changed{$G_{n,M}$} random graph~\cite{bollobas2001} of 7 nodes and density $\rho=0.68$ (14 edges). (b) Example of a \changed{random regular graph~\cite{bollobas2001}} of 7 nodes and degree $Z=4$ (14 edges). (c) The circuit of a single layer of the extended QAOA (see Eq.~\eqref{eqaoa}) (Color online).}
\label{Figure_graphs_QQAOA}
\end{figure}

\changed{Each problem's $J$ defines a weighted undirected graph with $N$ vertices, connected by edges $i \leftrightarrow j$ wherever $J_{ij},J_{ji}\neq0$. We study two families of random graphs~\cite{bollobas2001, suppl}. The \textit{$G_{n,M}$ random graphs} or  \textit{Erdös-Rényi graphs}, samples uniformly a predetermined number of edges $M$ (given by the density $\rho= 2M / (N^2-N) \in [0,1]$)  from all  possible connections [cf. Fig.~\ref{Figure_graphs_QQAOA}a]. In \textit{random $r$-regular graphs}, edges are randomly sampled, with the constraint that each vertex has $r$ neighbors [cf. Fig.~\ref{Figure_graphs_QQAOA}b].}

We create problems by sampling graphs with the Python package \textit{networkx}~\cite{networkx_2008,suppl_github} and randomly generating the Hamiltonian coefficients~\eqref{eq:energy}. In QUBO, the nonzero values of $Q_{ij}$ are randomly drawn from a normal distribution \changed{$N(\mu=0,\sigma^2)$ with zero mean $\mu$ and variance $\sigma^2=1$. In MaxCut, we also draw $Q_{ij}$ from $N(\mu=0,\sigma^2)$,} which includes the Sherrington-Kirkpatrick model~\cite{Sherrington_1975} in the fully connected case. Finally, in the random Ising model both $J$ and $h$ are generated with the same distribution \changed{$N(\mu=0,\sigma^2)$.}

\textit{Single-layer QAOA.}-- Let us introduce a generalization of the QAOA ansatz, consisting of a layer of Hadamard gates $H^{\otimes N}$, followed by evolution with the Hamiltonian $\hat{E}$ and single-qubit rotations $R_y^{\otimes N}$ and $U_1^{\otimes N}$, as shown in Fig.~\ref{Figure_graphs_QQAOA}c. The resultant state is given by:
\begin{equation}
\ket{\Psi(\gamma,\theta,\lambda)} \equiv R_y(\theta)^{\otimes N} U_1(\lambda)^{\otimes N} e^{-i\gamma \hat{E}} H^{\otimes N}\ket{0}^{\otimes N},
\label{eqaoa}
\end{equation}
with $R_y(\theta)\equiv \exp\left(-i\frac{\theta}{2}\sigma_y\right)$, $U_1(\lambda)\equiv\exp\left(-i\frac{\lambda}{2}\sigma_z\right)$, where $\sigma_{y,z}$ are the Pauli matrices. Here $\gamma,\theta >0$ are variational parameters. \changed{The phase $\lambda$ controls the optimization direction and $\lambda=\pm\frac{\pi}{2}$ produces a single layer QAOA~\cite{Farhi2014}.}

Circuit~\eqref{eqaoa} acts as an interferometer in energy space, amplifying the probability of states with high or low energy for $\lambda=\pi/2$ or $-\pi/2$, respectively. Indeed, for a single two-level system $\hat{E}=\frac{1}{2}\Delta\sigma^z$, the probability $P(s)$ to measure state $\ket{s=\pm 1}$ is
$P_s = \frac{1}{2}\left(1 - s \, \sin(\theta)\cos\left(\gamma \Delta + \lambda\right)\right)$. \changed{Under optimal angles $\theta=\frac{\pi}{2}$ and $\gamma=\frac{\pi}{2\Delta}$, the probability is maximized for the ground or excited state, depending on $\lambda=\pm \pi/2$.}

\begin{figure}
\includegraphics[width=1\linewidth]{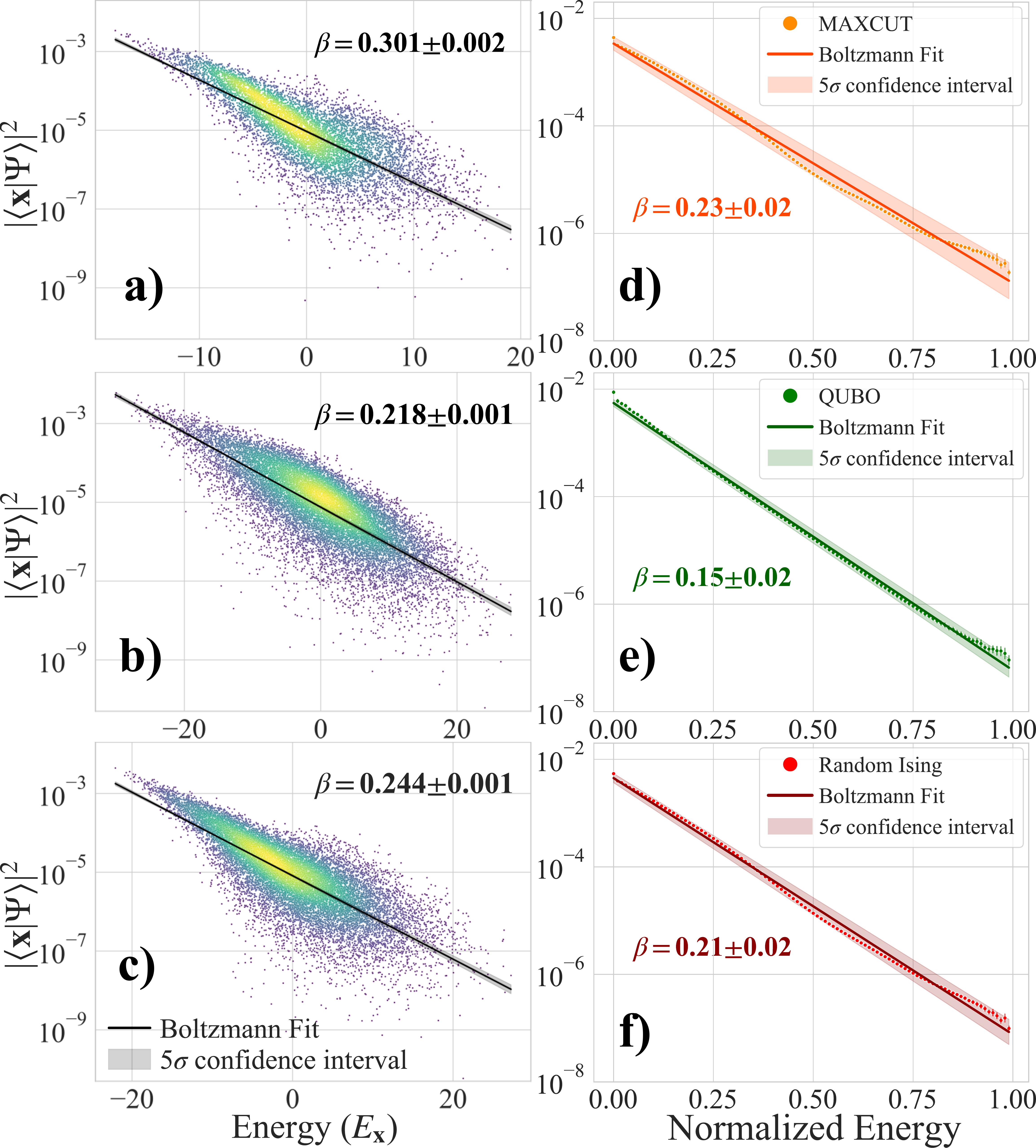}
\caption{Eigenstate probabilities (dots) and fitted Boltzmann distribution (line) for ansatz~\ref{eqaoa} with optimal angles in (a,d) Max-Cut, (b,e) QUBO and (c,f) Random Ising 14 qubits problems on \changed{$G_{n,M}$} random graphs with density equal to $0.9$. In (d,e,f) we normalized the energies and computed the probabilities (dots) over 100 energy intervals, averaged over \changed{1000} replicas. We show these values, together with a fit to a Boltzmann distribution. \changed{All results show the $99\%$ confidence interval in the fit (Color online).}}
\label{Figure_fits}
\end{figure}

\paragraph{Pseudo-Boltzmann states.--} We applied circuit~\eqref{eqaoa} to the variational minimization of QUBO, MaxCut and random Ising problems ($\lambda=-\pi/2$). \changed{We studied problems from 6 to 22 qubits, with varying graph density ($G_{n,M}$ random graphs) or coordination number (random regular graphs). Given these parameters, we generate randomly connected graphs, each with a different random matrix---$Q$, or $J$ and $h$ as explained above.}
For each problem, we \changed{searched the angles $\{\theta,\gamma\}$ that minimized the energy $\braket{\Psi(\gamma,\theta)|\hat{E}|\Psi(\gamma,\theta)}$}, computed the enhancement in ground state probability w.r.t uniform sampling, $\xi = \left|\braket{\mathbf{x}_0|\Psi}\right|^2 2^{N}$, and reconstructed the probability distribution $P(E) =\sum_\mathbf{x} \delta(E-E_\mathbf{x}) \left|\braket{\mathbf{x}|\Psi}\right|^2$ in energy space.

Figs.~\ref{Figure_fits}a-c show the optimized QAOA states for \changed{three random instances} with 14 qubits of (a) MaxCut, (b) QUBO and (c) random Ising problems. Even though they are pure states, the eigenstates' probabilities  $\left|\braket{\mathbf{x}|\Psi}\right|^2$ fluctuate around a straight line given by a Boltzmann distribution $\left|\braket{\mathbf{x}|\Psi}\right|^2 \sim \exp(-\beta E_x)$ with inverse temperature $\beta$. This is what we call a \textit{pseudo-Boltzmann} state. As shown in this figure, this fit has a high confidence and correlation, and the actual probabilities resemble $P(E)\sim \exp(-\beta E)\rho(E)$, with a density of states $\rho(E)$ similar to a normal distribution that \changed{concentrates on intermediate energies (see below).}

\changed{For further confirmation, we rescaled the eigenenergies of all 1000 replicas to the $[0,1]$ interval, binning them into 100 sub-intervals. This results in a replica-averaged probability distribution $\braket{P(E)}$ that is} fitted to a Boltzmann distribution, now independent of the Hamiltonian norm. Figs.~\ref{Figure_fits}d-f show three fits for the three families of random problems, for $N=14$ spins. \changed{The average inverse temperature $\beta$ displayed in these figures is estimated by dividing the slope of the fit, by the mean normalization factor $\braket{E_{max}-E_{min}}$ of the 1000 replicas}. This average $\beta$ is slightly lower than the one of individual instances, because of the energy normalization. We also notice that \changed{the QUBO models yield the most accurate fit.}

\begin{figure}
\includegraphics[width=1\linewidth]{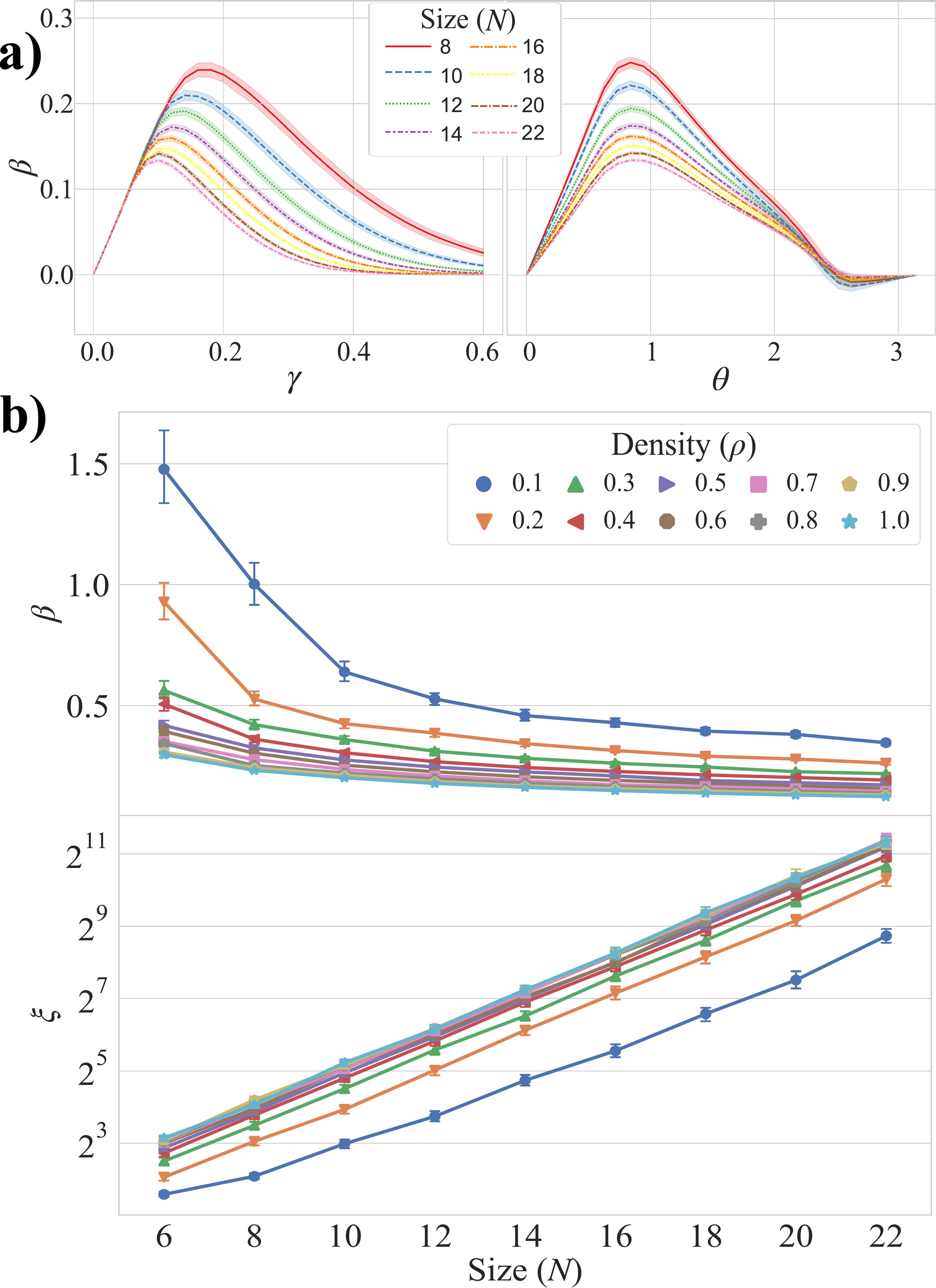}
\caption{(a) \changed{Average} effective temperature $\beta$ after one-layer QAOA for QUBO problems on \changed{$G_{n,M}$} random graphs with density equal to 0.9 and different number of nodes, as a function of the QAOA angles. When varying $\gamma$, we set $\theta$ to its optimal value \changed{for each individual instance}, and vice versa. (b) \changed{Average} effective temperature $\beta$ and ground state amplitude enhancement after one-layer QAOA with optimal angles for QUBO problems on \changed{$G_{n,M}$} random graphs. \changed{All results show the average over 500 instances with an error bar denoting the 99$\%$ confidence interval in the average estimation(Color online).} }
\label{Figure_beta_enhancement}
\end{figure}

\begin{table}[t]
\begin{tabular}{c|c|c|c|}
\cline{2-4}
\multicolumn{1}{l|}{}                                                                        & \multicolumn{1}{c|}{Optimal $\gamma$}                                             & \multicolumn{1}{c|}{$\xi$} & \multicolumn{1}{c|}{$\beta$}                                            \\ \hline
\multicolumn{1}{|c|}{\begin{tabular}[c]{@{}c@{}}$G_{n,M}$ Random \\ Graphs\end{tabular}}           & $ \dfrac{\chi_{\gamma}}{\sigma\sqrt{(N-1)\rho}}$  & $\sqrt{2^N}$     & $ \dfrac{\chi_{\beta}}{\sigma\sqrt{(N-1)\rho}}$  \\ \hline
\multicolumn{1}{|c|}{\begin{tabular}[c]{@{}c@{}}Regular \\ Graphs\end{tabular}}                       & $ \dfrac{\chi_{\gamma}}{\sigma\sqrt{Z}}$          & $\sqrt{2^N}$        & $ \dfrac{\chi_{\beta}}{\sigma\sqrt{Z}}$         \\ \hline
\multicolumn{1}{|c|}{\begin{tabular}[c]{@{}c@{}}General \\ Graphs\end{tabular}}             & $ \dfrac{\chi_{\gamma}\sqrt{N}}{\sigma\sqrt{2M}}$  & $\sqrt{2^N}$   & $ \dfrac{\chi_{\beta}\sqrt{N}}{\sigma\sqrt{2M}}$ \\ \hline
\end{tabular}
\caption{Approximated trends of the single-layer QAOA that we observe when $N$ is sufficiently large, as a function of the problem parameters~\cite{suppl}.  These results were obtained for three families of combinatorial optimization problems: MaxCut, QUBO, and random Ising models. \changed{The optimal $\theta$ converges to a constant value for MaxCut and QUBO models.} $\rho$, $Z$, and $M$ denotes the density, the degree, and the total number of edges of the graph respectively.\label{tab:scalings}}
\end{table}

\changed{Fig.~\ref{Figure_beta_enhancement} shows the QUBO temperature and the ground state enhancement as we change angles, problem size and the graph density, averaged over problems. For each instance we compute the angles $\{\theta_{M}, \gamma_{M}\}$ that
minimize the cost function and then fix $\theta=\theta_M$ and vary $\gamma$ (Fig.~\ref{Figure_beta_enhancement}a, left), or set $\gamma=\gamma_M$ and change $\theta$ (right). In both cases, there is a maximum at which the interferometer wraps around and the Boltzmann distribution breaks down. This happens around $\theta\simeq \pi/3$ and $\gamma\sim \mathcal{O}(N^{-1/2})$ for highly connected graphs, which is consistent with the scalings of $\theta_M$ and $\gamma_M$ found in Ref.~\cite{Ozaeta2020}. The inverse temperature [cf. Fig.~\ref{Figure_beta_enhancement}b] is large enough and decreases slowly with problem size $\beta\sim \mathcal{O}(N^{-1/2})$, enabling an exponential enhancement of the ground state probability $\xi\sim \mathcal{O}(2^{N/2})$ [cf. Fig.~\ref{Figure_beta_enhancement}c]. Table~\ref{tab:scalings} describes how these results generalize to other graphs and problems.}

\paragraph{Derivation of pseudo-Boltzmann states.--}%
\changed{QAOA circuits~\eqref{eqaoa} create} a uniform superposition of all problem eigenstates $\mathbf{x}$, which evolve for a time $\gamma$ with their energies $E_\mathbf{x}$ and later interfere in the local rotations $\lambda$ and $\theta$. The amplitude of final states $F(\mathbf{x})= \braket{\mathbf{x}|\Psi}$ \changed{is~\cite{suppl}}
\begin{equation}
  F(\mathbf{x}) \propto \braket{\exp\left({H_{\mathbf{x}\mathbf{x}'}(-i\lambda - r_\theta)-i\gamma E_{\mathbf{x}'}}\right)}_{\mathbf{x}'}.
\end{equation}
Here $r_\theta=-\log(\tan(\theta/2))$, $H_{\mathbf{x}\mathbf{x}'}$ is the Hamming distance between two bit strings, and we average uniformly over all spin configurations $\mathbf{x}'$.

\changed{For large problems, the average over configurations becomes an average over a quasi-continuous probability distribution $p(H, E, \mathbf{x})$, combining the eigenenergies $E$ and the Hamming distances $H$ from eigenstates to the reference $\mathbf{x}$. For our models, $p(H, E, \mathbf{x})$ approaches a sum over one or two Gaussians, depending on the symmetries and degeneracy of the problem~\cite{suppl}. The Gaussians
\begin{equation}
  p(H,E,\mathbf{x}) \propto e^{-\frac{1}{2(1-\rho^2)}\left[\left(\frac{E}{\sigma_E}\right)^2
      + \left(\frac{H-\mu_H}{\sigma_H}\right)^2
      - 2 \rho \frac{E(H-\mu_H)}{\sigma_E\sigma_H}\right]},
\label{gaussian}
\end{equation}
are centered on $\mu_E=0$ and $\mu_H$, with standard deviations $\sigma_E$ and $\sigma_H$. The correlation factor $ \rho(\mathbf{x}) = (\sigma_E\sigma_H)^{-1}\sigma_{EH}(\mathbf{x})$,
is the covariance $\sigma_{EH}(\mathbf{x})$. Empirically, the small $\rho(\mathbf{x})$ dominates the integral in $F(\mathbf{x})$, so that
$\vert{F(\mathbf{x})}\vert^2 \propto e^{-2\gamma \lambda \sigma_{EH}(\mathbf{x})},$
up to normalization~\cite{suppl}.}

\begin{figure}[t]
  \centering
\includegraphics[width=\linewidth]{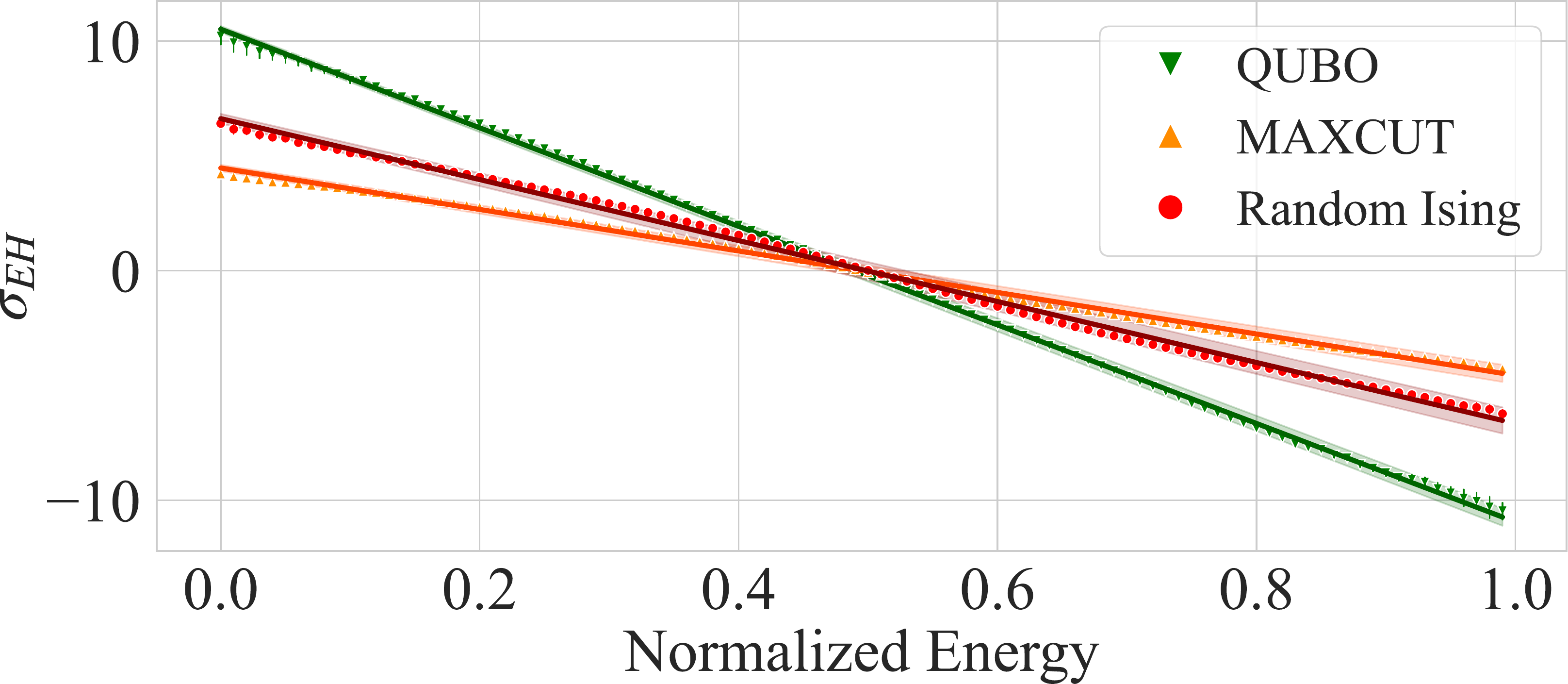}
\caption{Average covariance $\sigma_{EH}$ as a function of the normalized energy of the state $E_\mathbf{x}$ over 300 replicas for QUBO, MaxCut and random Ising problems on \changed{$G_{n,M}$} random graphs with 14 nodes \changed{and density 0.9}. Note that $\sigma_{EH}\equiv\sigma_{EH+}$ for degenerate problems, as explained in \cite{suppl}. \changed{We show the 99$\%$ confidence area of the linear regression (Color online).}}
  \label{fig_correlation}
\end{figure}

\changed{We estimate $\sigma_{EH}(\mathbf{x})$ for each problem by sampling all pairs $(H_{\mathbf{x},\mathbf{x}'},E_{\mathbf{x}'})$ for all eigenstates $(\mathbf{x},\mathbf{x}')$, after rescaling the energies as in Fig.~\ref{Figure_fits}d-f. Using a variational Bayesian estimation method~\cite{scikit-learn}, we fit these samples to one or more Gaussians~\eqref{gaussian}. Fig.~\ref{fig_correlation} displays the binned outcomes from multiple fits, expressing $\sigma_{EH}(\tilde{E}_\mathbf{x})$ as a function of the rescaled energies. We conclude that $\sigma_{EH}(\mathbf{x})\sim-c E_\mathbf{x}+\omega$ with small random fluctuations $\omega$.} This allows us to write
\begin{equation}
  \vert{F(\mathbf{x})}\vert^2 \propto \exp(-\beta E_\mathbf{x} + \beta \omega/c),
\end{equation}
with the effective temperature $\beta = -2c\gamma \lambda$. \changed{Intrinsic degeneracies, as in MaxCut, induce corrections to all formulas which do not affect the prediction~\cite{suppl}. Finally, as additional evidence that $p(E,H,\textbf{x})$ explains the pseudo-Boltzmann state we verified that when we remove all structure, assigning random values to $E_\textbf{x}$, we recover uniform distributions with $\beta\to0$ (not shown here).}

\paragraph{\changedsecondround{Pseudo-Boltzmann states sampling hardness.--}}%
It has been shown that single-layer QAOA circuits have a quantum advantage, in the sense that they are classically hard to sample~\cite{Farhi2019}, for if they were not, the polynomial hierarchy would collapse. This result is related to the conjecture that the complex partition function of an Ising model, recreated by the circuit, is hard to sample~\cite{Bremner_2016}.

\changed{Our work suggests that the hardness of sampling single-layer QAOA circuits is a consequence of the low effective temperature of the pseudo-Boltzmann states, which scales favorably with the problem size [cf. Table~\ref{tab:scalings}].} \changedsecondround{There is an active field of research on classical algorithms capable of efficiently sampling from the Boltzmann distribution of Ising models at a given temperature $\beta<\beta_T$, where $\beta_T$ is a threshold that ensures rapid mixing}. \changedsecondround{Following the approach of previous works in the literature~\cite{GarciaPatron2021}, we compare the effective temperature reached by single-layer QAOA states with the state-of-the-art general condition that warrants fast mixing for Ising models, $\beta_\text{MCMC} < \Vert{J}\Vert^{-1}$~\cite{eldan2021spectral}.}

\begin{figure}
  \centering
\includegraphics[width=\linewidth]{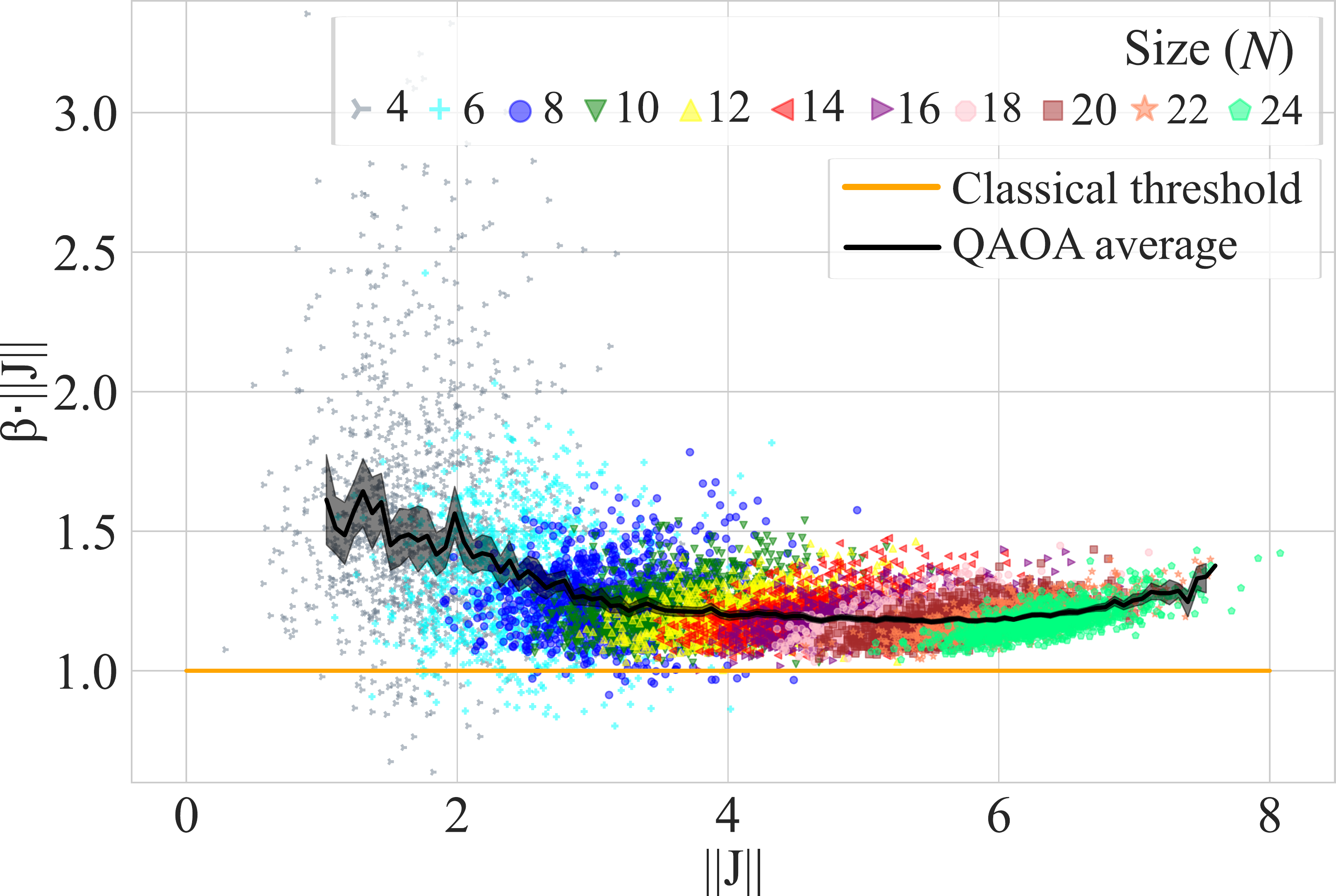}
\caption{Effective temperatures $\beta$ after single-layer QAOA with optimal angles, as a function of the interaction matrix \changed{operator} norm $\Vert{J}\Vert$ for \changed{the Sherrington-Kirkpatrick model}. The dots display the results for \changed{1000} instances for each problem size, which grows with $||J||$. We show the average of these results (black line) and the threshold given by Monte Carlo methods (orange line) (Color online).}
  \label{Figure_quantumadvantage}
\end{figure}

\changed{Indeed, }\changedsecondround{ we find that,} \changed{for some models such as the Sherrington-Kirkpatrick, the inverse temperature of the QAOA state grows beyond} \changedsecondround{this regime in which rapid mixing asserts the efficient sampling of the distribution by Markov Chain Monte Carlo (MCMC) methods ~\cite{eldan2021spectral}.} \changed{As illustration, Fig.~\ref{Figure_quantumadvantage} shows the QAOA temperature for 11000 instances of the SK model for up to 24 qubits, as a function of the interaction matrix operator norm $\Vert{J}\Vert$. Note the stable gap between $\beta_\text{QAOA}\Vert{J}\Vert$ and the upper bound of the rapid mixing regime for MCMC, which scales as $\beta_\text{MCMC}= \Vert{J}\Vert^{-1} < (2\sqrt{N})^{-1}$ when $J$ is sampled according to $\mathcal{N}(0,1)$~\cite{Anderson2009}.}
\changedsecondround{These results provide further insight into previous theoretical results on the difficulty of sampling QAOA~\cite{Farhi2019}. It is feasible that the enhancement of the ansatz expressibility produced by increasing the number of circuit layers has an impact on a rise in the effective inverse temperature of the QAOA states, provided that the pseudo-Boltzmann distribution of these states is maintained. We leave this question for future work. Separately, although these results may have implications for optimization tasks, it is not yet understood whether they translate into a practical advantage when compared to state-of-the-art methods, such as replica exchange sampling. We also leave this point for further study.}

\paragraph{Summary and discussion.--}%
\changed{Motivated by their formal advantages~\cite{Farhi2019}, we have investigated the states generated by single-layer QAOA circuits. Our work focused on NP-hard problem Hamiltonians on random graphs of different type and dimensions. The simulations and the accompanying theory reveal that single-layer QAOA circuits sample the eigenstates of a problem Hamiltonian according to a Boltzmann distribution with fluctuations. When averaged over replicas, the effective temperature of this \textit{pseudo-Boltzmann} pure state scales favorably with the problem size}\changedsecondround{, enhancing the probability of finding the ground state by an algebraic factor over uniform sampling $\xi\sim \mathcal{O}(2^{N/2})$, similar to Grover-type search algorithms}. The results in this work are built upon a qualitative understanding of the structure of eigenstates and their correlations in Ising-like spin models. This analysis merits further investigation, to understand whether the correlations that support QAOA's \changed{behaviour}
may translate to other problems.

\changed{
Despite similarities, the pseudo-Boltzmann states are unrelated to the eigenstate thermalization hypothesis (ETH, see Ref.~\cite{D_Alessio_2016}). ETH correlations exhibit thermal-like behavior for long times, but single-layer QAOA pseudo-Boltzmann states appear only after a very short evolution $\mathcal{O}(\gamma)$ [c.f. Fig.~\ref{Figure_beta_enhancement}a], and acquires an infinite temperature $(\beta\to 0)$ for long times. Moreover, ETH systems exhibits a thermal-like behavior with respect to the Hamiltonian $H$ that describes the evolution. In contrast, QAOA dynamics is governed by a combination $H\sim \gamma \times \hat{E} + \theta H_\text{field}$ that includes a transverse field $H_\text{field}\sim \sum_n \sigma_y^{(i)}$, but the final state is thermal with respect to just the problem Hamiltonian $\hat{E}$. Finally, unlike ETH, the QAOA thermalization cannot be explained by a chaotic energy spectrum, requiring the hidden energy-distance correlation discussed above.}

\changed{We believe that this work may have deeper implications in the study of both quantum simulators and more complex variational algorithms. The single-layer QAOA evolution that we have used in this work is equivalent to a short-time evolution with an effective Hamiltonian that combines Ising interactions $\hat{E}$ and transverse field $H_\text{field}$. Not only can this physics be explored in existing quantum simulators, but it is also relevant to understand whether similar pseudo-Boltzmann states can be produced by more general time-dependent controls of the interaction and the mixing terms, $H(t) = s(t) \hat{E} + g(t) H_\text{field}$. Eventually, one would expect that this line of research could explain both the limit of perfect adiabatic passages $\beta\to+\infty$ as well as the rate of growth of imperfections in a quantum annealer.}

\changed{Finally, we believe that the appearance of single-layer pseudo-thermal states explains the success in approximating thermal states using multi-layer QAOA-like variational mixed-state ansätze with entropy sources~\cite{verdon2019quantum}. While it is unclear how to generalize the interferometric picture to a multi-layer scenario, one may still analyze the creation of pseudo-Boltzmann states in presence of multiple layers, without the need of the purification or entropy source~\cite{verdon2019quantum}. This line of work may also provide a useful connection to the the investigation of annealing passages $s(t), g(t)$ suggested above.}

The supporting data for this letter are openly available
from~\cite{suppl_github}.

\begin{acknowledgments}
We acknowledge Centro de Supercomputación de Galicia (CESGA) who provided access to the supercomputer FinisTerrae for performing simulations. We thank Fernando J. Gómez-Ruiz for critical reading of the manuscript. This work has been supported by European Commission FET Open project AVaQus Grant Agreement 899561, PGC2018-094792-BI00 (MCIU/AEI/FEDER,UE), Comunidad de Madrid Sinergicos 2020 project NanoQuCo-CM (Y2020/TCS-6545), CSIC Quantum Technologies Platform PTI-001, and CAM/FEDER project No. S2018/TCS-4342 (QUITEMAD-CM).
\end{acknowledgments}

\bibliographystyle{apsrev4-2}
\bibliography{bibliography}

\clearpage
\onecolumngrid

\begin{center}
  {\large\textbf{Supplementary material to ``Quantum Approximate Optimization Algorithm pseudo-Boltzmann states''}}
\end{center}

\begin{center}
  P. Díez-Valle$^1$, D. Porras$^1$, J. J. García Ripoll$^1$

  \textit{$^1$Instituto de F\'{\i}sica Fundamental IFF-CSIC, Calle Serrano 113b, Madrid 28006, Spain}\\
\end{center}
\vspace{0.5cm}

In this document we complete the discussion from our work on ``QAOA pseudo-Boltzman states'', detailing the mathematical proofs of the emergence of pseudo-Botlzmann state, providing a more thorough investigation of the underlying correlations that support those states, and illustrating the results from the numerical simulations whose aggregate results we aggregate into the various average plots and scalings.
\\

\section{I. Theoretical derivations}

\subsection{A. QAOA interference amplitude}

The amplitudes of the state generated by the single-layer QAOA circuit are given by the interference formula
\begin{equation}
F(\mathbf{x}) \equiv \braket{\mathbf{x}|\tilde\Psi} = \frac{1}{2^{N/2}}\sum_{\mathbf{x'}} \cos\left(\frac{\theta}{2}\right)^{N-H_{\mathbf{x},\mathbf{x'}}}\left[e^{-i\lambda} \sin\left(\frac{\theta}{2}\right)\right]^{H_{\mathbf{x},\mathbf{x'}}}
e^{-i\gamma E_{\mathbf{x'}}}.
\end{equation}
In this equation $H_{\mathbf{x},\mathbf{x'}}$ is the Hamming distance between two configurations of bits $\mathbf{x}$ and $\mathbf{x'}$ which are eigenstates of the spin model whose energy is denoted $E(\mathbf{x})$ in Eq.~(1) from the manuscript, and here abbreviated as $E_{\mathbf{x'}}$.

The weights in the sum are expressed in terms of a rotation angle $\theta$ which can be reparameterized as one exponent $r$ and a normalization $R$
\begin{equation}
\cos\left(\frac{\theta}{2}\right) = R^{\frac{1}{2}} \exp\left(\frac{r}{2}\right), \; \sin\left(\frac{\theta}{2}\right) = R^{\frac{1}{2}}\exp\left(-\frac{r}{2}\right).
\label{eq_cosin}
\end{equation}
With this notation, the amplitudes become
\begin{equation}
F(\mathbf{x}) = \left(\frac{R\exp (r)}{2}\right)^{\frac{N}{2}}\sum_{\mathbf{x'}}\exp\left[H_{\mathbf{x}\mathbf{x'}}(-i\lambda - r) - i\gamma E_{\mathbf{x'}}\right].
\end{equation}

The previous sum runs over eigenstates of the spin model $\textbf{x}$. We can introduce a probability distribution
\begin{equation}
  p(H, E; \mathbf{x})
   = \frac{1}{2^{N}}\sum_{\mathbf{x'}} \delta(H - H_\mathbf{xx'}) \delta(E - E_\mathbf{x'}),
\end{equation}
that captures the internal correlations and degeneracies of the particular spin model. This allows us to rewrite the interference amplitude as an average over this probability distribution
\begin{equation}
  F(\mathbf{x}) \propto \left(\exp (r)\right)^{\frac{N}{2}}\int\int_{-\infty}^{\infty}
  \exp\left[H(-i\lambda - r) - i\gamma E\right] p(H,E; \mathbf{x}) dHdE.
  \label{eq_amplint}
\end{equation}

\begin{figure}
\includegraphics[width=0.5\linewidth]{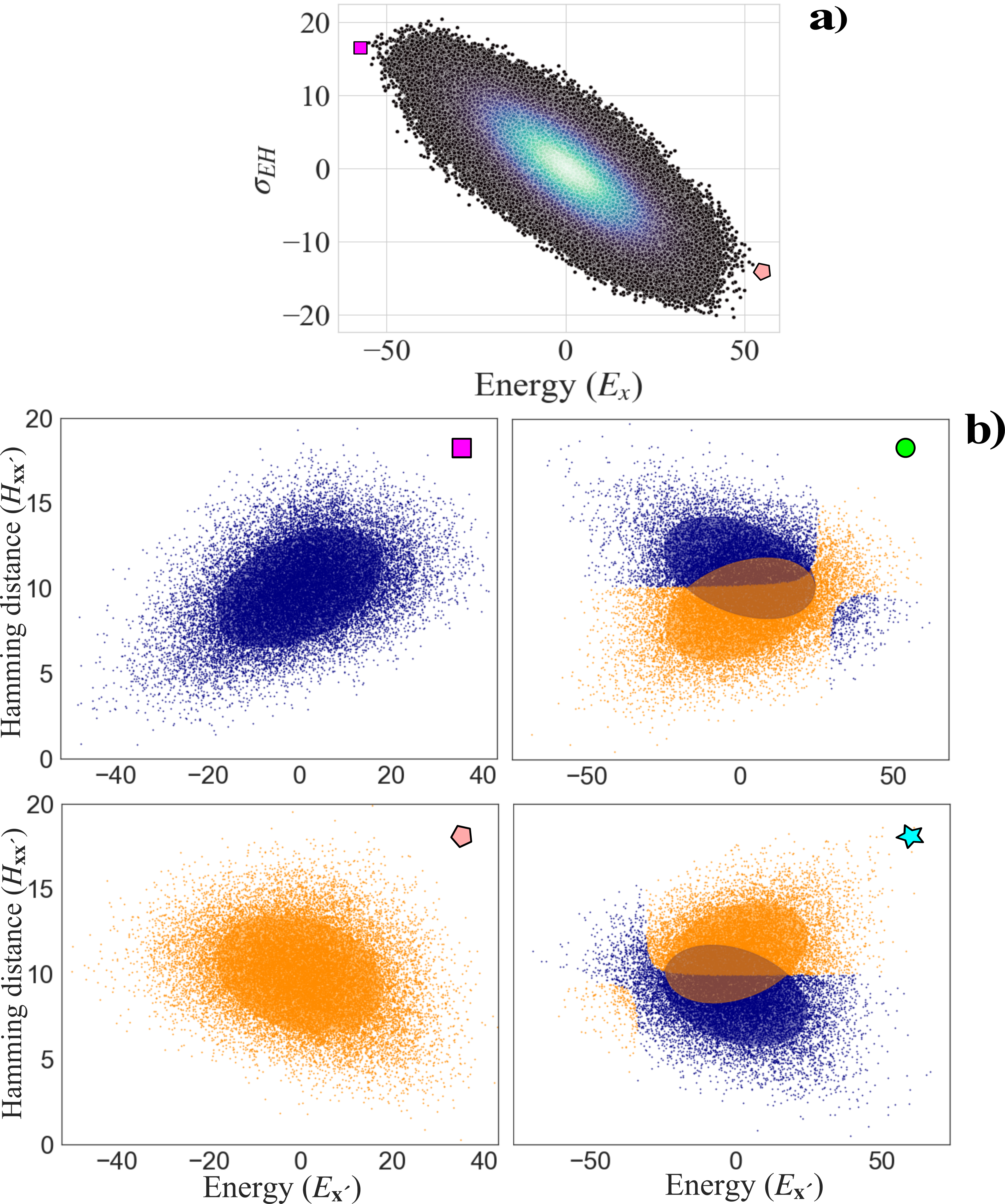}
\caption{(a) Correlation between the covariance $\sigma_{EH}$ and energy $E_\mathbf{x}$ for one QUBO instance on Gnm random graph with 20 nodes. (b) Continuous probability distribution $p(H,E; \mathbf{x})$ estimated by 25000 samples drawn from a kernel density estimation (KDE). We show the distributions for one instance of (square and pentagon) non-degenerate QUBO and (circle and star) degenerate MaxCut model, where $\mathbf{x}$ is (square and circle) the ground state, and (pentagon and star) the highest energy state of the system. Over these probability distributions, we plot the confidence ellipsoids of the fit to a Gaussian mixture obtained with Variational Inference. In specific, the fit was obtained with the \textit{BayesianGaussianMixture} method of the Python package \textit{scikit-learn} (Color online).}
\label{Fig:covariance_distributions}
\end{figure}

In our study of the NP-hard models for random QUBO, random MaxCut and random Ising we have found that the probability distribution $p(H, E;\mathbf{x})$ exhibits slightly different behaviors depending on whether the model has a global $\mathbb{Z}^2$ symmetry or not. In the latter case $p(H,E;\mathbf{x})$ becomes a single Gaussian distribution [cf. Fig.~\ref{Fig:covariance_distributions}b(left)], while in the former one it may be a sum of slightly displaced probability distributions [cf. Fig.~\ref{Fig:covariance_distributions}b(right)]. We now discuss this casuistic.

\subsection{B. Non-degenerate model probability distribution}

In many of the models we have studied, such as the random QUBO problem, there exist no global symmetries in the problem. In those situations the eigenstates seem to order themselves from ground to excited states, developing a single probability distribution $p(H,E;\mathbf{x})$ that is centered in the middle of the spectrum and exhibits a changing correlation between $H_\mathbf{xx'}$ and $E_\mathbf{x'}$.

This is best seen in Fig.~\ref{Fig:covariance_distributions}b. Take for instance the upper-left plot from, which shows the probability distribution $p(H,E;\mathbf{x})$ when  $\mathbf{x}$ is the ground state. All other states $\mathbf{x'}$ have a higher energy than $E_\mathbf{x}$. The difference $E_\mathbf{x'}-E_\mathbf{x}$ may be correlated to the number of times a spin must be flipped to go from $\mathbf{x}$ to $\mathbf{x'}$ (i.e. the Hamming distance $H_\mathbf{xx'}$). This positive correlation manifests in a Gaussian function that is oriented along a diagonal. Conversely, if we choose $\mathbf{x}$ to be the highest energy state, we find the opposite trend: the more spins we flip, the lower the energy.

In this non-degenerate situation the probability distribution $p(H,E; \mathbf{x})$ can be well approximated by a bivariate Gaussian where the covariance $\sigma_{EH}$ depends on $\mathbf{x}$
\begin{equation}
  p(H,E; \mathbf{x})_{\rho}\approx
  \dfrac{1}{2\pi\sigma_E\sigma_H\sqrt{1-\rho^{2}}}
  \exp\left[-\dfrac{1}{2(1-\rho^{2})}
    \left(\left(\dfrac{E}{\sigma_E}\right)^2+\left(\dfrac{H-\mu_H}{\sigma_H}\right)^2-2\rho\dfrac{E(H-\mu_H)}{\sigma_E\sigma_H}\right)\right],
  \label{eq_bigaussian}
\end{equation}
where
\begin{equation}
\mu_E=0\,;\,\mu_H=\frac{N}{2}\,;\,\sigma_H= \frac{\sqrt{N}}{2}\,;\,\rho = \dfrac{\sigma_{EH}(\mathbf{x})}{\sigma_E\sigma_H}.
\end{equation}

Note that the correlation parameter $\rho$ encapsulates the dependence in $\mathbf{x}$ since the features of the spin model fix the rest of the values. Then, Eq.~\eqref{eq_amplint} and Eq.~\eqref{eq_bigaussian} lead to
\begin{align}
    |F(\mathbf{x})|^2 &\propto \exp[Y],\mbox{ with}
    \label{eq_anatampl}\\
    Y &= -\gamma ^2 \sigma_E^2
    +\left(r^2-\lambda^2\right)\sigma_H^2-2r\mu_H-2 \gamma  \lambda \rho \sigma_E \sigma_H .
   \label{eq_yaprox}
\end{align}
Connecting with the discussion above, in the simulations we find that $\rho(\mathbf{x})$ is approximately linear, decreasing from the largest, most positive value $\rho(\mathbf{x})>0$ when $\mathbf{x}$ is the ground state, down to the smallest, most negative value $\rho(\mathbf{x})<0$ when the reference $\mathbf{x}$ is the highest energy state.

\subsection{C. Degenerate model probability distribution}

When the Hamiltonian exhibits some global symmetries, the Hilbert space may split into two or more hierarchies of eigenstates which begin from different ground states. Among the problems we study, the random MaxCut exhibits a $\mathbb{Z}_2$ symmetry that leads to such a degeneracy.

In presence of this global symmetry, if $\mathbf{x}^{(a)}=(x_1^{(a)},x_2^{(a)},\ldots )$ is a ground state, then we have also another ground state in the opposite sector ${\mathbf{x}}^{(b)}=1-{\mathbf{x}^{(a)}}$. We may now regard a single excited state $\mathbf{x'}$ as the result of flipping spins in either of the ground states $\mathbf{x}^{(a)}$ or $\mathbf{x}^{(b)}$. The result is that the same spin configuration has macroscopically different Hamming distances to both ground states, but still has the same energy. In particular, note that the separation between the two degenerate ground states $H_{\mathbf{x}^{(a)}\mathbf{x}^{(b)}} = N$ saturates the Hamming distance.

This phenomenon leads to a separation of the probability distribution as a sum of two probabilities, one measured with respect to each of the hierarchy of states
\begin{equation}
   p(H,E; \mathbf{x}) = p_{+}(H_{\mathbf{x},\mathbf{x'}},E_{\mathbf{x'}};\mathbf{x}) + p_{-}(H_{\mathbf{x},\mathbf{x'}},E_{\mathbf{x'}};\mathbf{x}).
   \label{eq_bigaussian_plusminus}
\end{equation}
We have found that these distributions can be well fitted by two shifted bivariate Gaussian distributions (see Fig. \ref{Fig:covariance_distributions}):
\begin{align}
  p_{+}(H,E;\mathbf{x})_{\rho_{+}}&\approx
  \dfrac{1}{2\pi\sigma_E\sigma_H\sqrt{1-\rho_{+}^{2}}}
  \exp\left[-\dfrac{1}{2(1-\rho_{+}^{2})}
    \left(\left(\dfrac{E}{\sigma_E}\right)^2+\left(\dfrac{H-\mu_{H}+h_0}{\sigma_H}\right)^2-2\rho_{+}\dfrac{E(H-\mu_{H}+h_0)}{\sigma_E\sigma_H}\right)\right],
  \label{eq_bigaussian_plus}\\
  p_{-}(H,E;\mathbf{x})&\approx p_{+}(-H+2\mu_H,E;\mathbf{x})_{\rho_{-}},
  \label{eq_bigaussian_minus}
\end{align}
where $\mu_H=N/2$, $h_0>0$ is a constant shift and we have two different correlation factors $\rho_{\pm}(\mathbf{x}) = \dfrac{\sigma^{\pm}_{EH}(\mathbf{x})}{\sigma_E\sigma_H}$.

Once more, the factor $\rho_{\pm}$ encapsulates all dependence in $\mathbf{x}$, but now we have two functions that interfere with each other
\begin{align}
    |F(\mathbf{x})|^2 &\propto \exp[Y']\cdot\left(\cos{\left[2h_{0}\lambda+r\gamma(\rho_{+}+\rho_{-})\sigma_E\sigma_H\right]}+\cosh{\left[2h_{0}r-\gamma\lambda(\rho_{+}+\rho_{-})\sigma_E\sigma_H\right]}\right),\mbox{ with}
    \label{eq_anatampl_degenerate}\\
    Y' &\equiv -\gamma ^2 \sigma_E^2+\left(r^2-\lambda^2\right)\sigma_H^2-2r\mu_H+ \gamma  \lambda  (\rho_{-}-\rho_{+})  \sigma_E \sigma_H
\end{align}
fortunately, since $|\rho_{\pm}|\sim \rho \ll 1$ for most states, the $h_0$-shift makes one exponential dominant
\begin{equation}
\begin{split}
       2\cosh{\left[2h_{0}r-\gamma\lambda(\rho_{+}+\rho_{-})\sigma_E\sigma_H\right]} = & \exp{\left[\beta'(\rho_{+}+\rho_{-})+2h_{0}r\right]} + \exp{\left[-\beta'(\rho_{+}+\rho_{-})-2h_{0}r\right]} \\
       \approx & \exp{\left[\beta'(\rho_{+}+\rho_{-})+2h_{0}r\right]},
    \label{cosh_approx}
\end{split}
\end{equation}
with $\beta'\equiv-\gamma\lambda\sigma_E\sigma_H$. This brings the amplitude to a formula
\begin{align}
    |F(\mathbf{x})|^2 &\propto \exp{[Y']}\cdot\exp{\left[\beta'(\rho_{+}+\rho_{-})\right]} = \exp[Y]
    \label{eq_anatampl_degenerate2},\mbox{ with}\\
    Y &= -\gamma ^2 \sigma_E^2
    +\left(r^2-\lambda^2\right)\sigma_H^2-2r\mu_H-2 \gamma  \lambda \rho \sigma_E \sigma_H ,
   \label{eq_yaprox_deg}
\end{align}
that is compatible with the non-degenerate scenario~\eqref{eq_anatampl}, except for small corrections due to the combination of eigenstate hierarchies~\eqref{cosh_approx}.

\subsection{D. Pseudo-Boltzmann states}

To study the probability amplitude distribution of the energy states, we can focus on the the terms that depend on the spin configuration $\mathbf{x}$ and its associated energy. Out of all contributions to $Y$, the only significant part is
\begin{equation}
Y \approx -2\gamma\lambda \rho_{EH} \sigma_E \sigma_H  = -2\gamma \lambda \sigma_{EH}(\mathbf{x}),
\end{equation}
in both the non-degenerate and degenerate models. This allows us to express the quantum probability distribution as an exponential
\begin{equation}
|F(\mathbf{x})|^2 \propto \exp(-2\gamma \lambda \sigma_{EH}(\mathbf{x})).
\label{eq_amplcov}
\end{equation}

Let us recall that $\sigma_{EH}(\mathbf{x})$ is the convariance between the energy of excited states and their Hamming distance to the reference $\mathbf{x}$. We have numerically observed a clear correlation between $\sigma_{EH}$ and the energy of the state $\mathbf{x}$, illustrated in Figs.~\ref{fig_correlation} and \ref{Fig:covariance_distributions}. This approximate dependency can be written as the sum of a linear function with positive slope $c>0$, and a stochastic value $\omega$ with zero mean
\begin{equation}
    \sigma_{EH}(\mathbf{x}) = -c\cdot E_{\mathbf{x}} \pm\omega.
    \label{eq_covEcorr}
\end{equation}
Despite the presence of the random term $\omega$, the tendency \eqref{eq_covEcorr} is manifest, and this correlation holds for all the studied families of Ising models. Therefore, from Eq.~\eqref{eq_amplcov} we get to express \eqref{eqaoa} as a thermal-like state:
\begin{equation}
|F(\mathbf{x})|^2 \propto \exp(-\beta E_\mathbf{x}\pm  \beta \omega/c),
\label{eq_amplasbolt}
\end{equation}

\begin{figure}[h]
\includegraphics[width=1\linewidth]{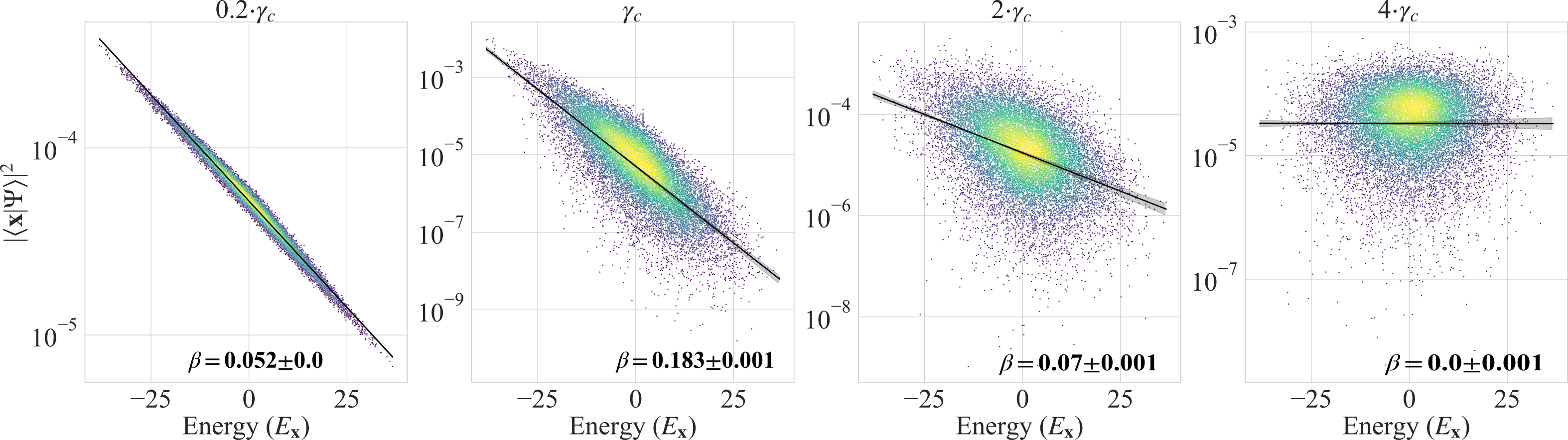}
\caption{Evolution of the eigenstates probabilities (dots) and fitted Boltzmann distribution (line) as we manipulate the angle $\gamma$. The plots show an individual random instance of a QUBO problem on a $G_{n,M}$ random graph with 14 nodes and density 0.9, for which we calculated the optimal QAOA angles $\{\gamma_{M},\theta_{M}\}$. With $\lambda = -\frac{\pi}{2}$ and $\theta=\theta_{M}$, we show the probability distribution of the eigenstates $\mathbf{x}$ after the single-layer QAOA when $\gamma$ is equal to $0.2\gamma_c$, $\gamma_c$, $2\gamma_c$, and $4\gamma_c$, where $\gamma_c = \gamma_{M}$ (Color online).}
\label{Fig:prob_gammachange}
\end{figure}

The overall Boltzmann fit combined with the random fluctuations $\omega$ are clear in the regimes of optimized parameters [cf. Fig.~\ref{Figure_fits}]. However, the distribution may be slightly transformed by manipulating the angles $\lambda$ and $\gamma$ [cf. Fig.~\ref{Fig:prob_gammachange}].
\begin{itemize}
    \item When $-\lambda=\frac{\pi}{2}$ we obtain a thermal state with temperature $T=\beta^{-1}=\frac{1}{c\pi\gamma}$ so that the maximum amplitude corresponds to the ground state of the system.

    \item Changing the sign of this angle to $\lambda=\frac{\pi}{2}$ is equivalent to changing the sign of the temperature. These negative temperature states amplify the probability of the highest excited state, instead of the ground state.

    \item We recover pseudo-Botlzmann states for a finite range of angles $\gamma \in (0,\gamma_c]$ ($t\in\mathbb{R}$, $\lambda >0$). The lowest temperature $T$ (highest $\beta$) is reached close to $\gamma \approx \gamma_c$, which is the optimal angle in the QAOA variational principle.

    \item When $\gamma > \gamma_c$, in the non-degenerate scenario every term in Eq.~\eqref{eq_anatampl} becomes negligible compared to $-\gamma^2\sigma_E^2$ and $|F(\mathbf{x})|^2\approx cte$, and in the degenerate cases the approximation~\eqref{cosh_approx} also becomes inaccurate.

    \item For smaller $\gamma < \gamma_c$ we find that $\beta$ grows approximately linearly with the angle $\gamma$, consistently with the predictions above.
\end{itemize}

\section{II. Simulations}

\changed{The code to reproduce all results shown in the manuscript is available as supplementary material in a Zenodo/GitHub repository~\cite{suppl_github}.}

\subsection{A. Problem generation}

In this work, we analyzed two well-known random graph structures to embed the optimization problems, i.e to define the non-zero elements of the interaction matrices $Q$ and $J$. A random graph starts without any edge, and it is built adding successive edges between the nodes with a procedure that satisfies some requirement given for the specific model. One model that we use is the so-called $G_{n,M}$ \textit{random graph}, one of the variants of the Erdös Renyi model in which a graph is chosen uniformly at random among all possible graphs with $n$ nodes and $M$ edges. Therefore, the probability assigned to every graph with those features is 1 divided by $\begin{pmatrix}n\\M
\end{pmatrix}$, where $n>M$ and $\begin{pmatrix}n\\M
\end{pmatrix}$ is the number of lattices that fulfill those conditions. We generated the adjacency matrix of this set of graphs from a fixed density $\rho$ with which the number of edges is calculated as $M = \textit{ceiling}\left[\rho(N^2-N)/2\right]$. The other lattice structure in our study is known as \textit{random r-regular graph}, and it is defined by the fact that every node has exactly $r$ neighbors. The number of neighbors $r$ is also called coordination number or degree $Z$. The total number of edges in a r-regular graph is given by $r\cdot n/2$, where n is the number of vertices, and therefore $r\cdot n$ must be even.

In practice, all the graphs were randomly generated by the python package \textit{networkx}~\cite{networkx_2008}. Each graph defines a connectivity matrix, where the non-zero elements are then drawn from a normal distribution such that $Q_{i,j},J_{i,j}\sim N(\mu=0,\sigma^2)$ for $i,j$ neighbors, with $\sigma^2=1$ unless otherwise stated, and otherwise $Q_{i,j},J_{i,j} = 0$.

\subsection{B. Approximated trends}

\begin{figure}[h]
\includegraphics[width=0.9\linewidth]{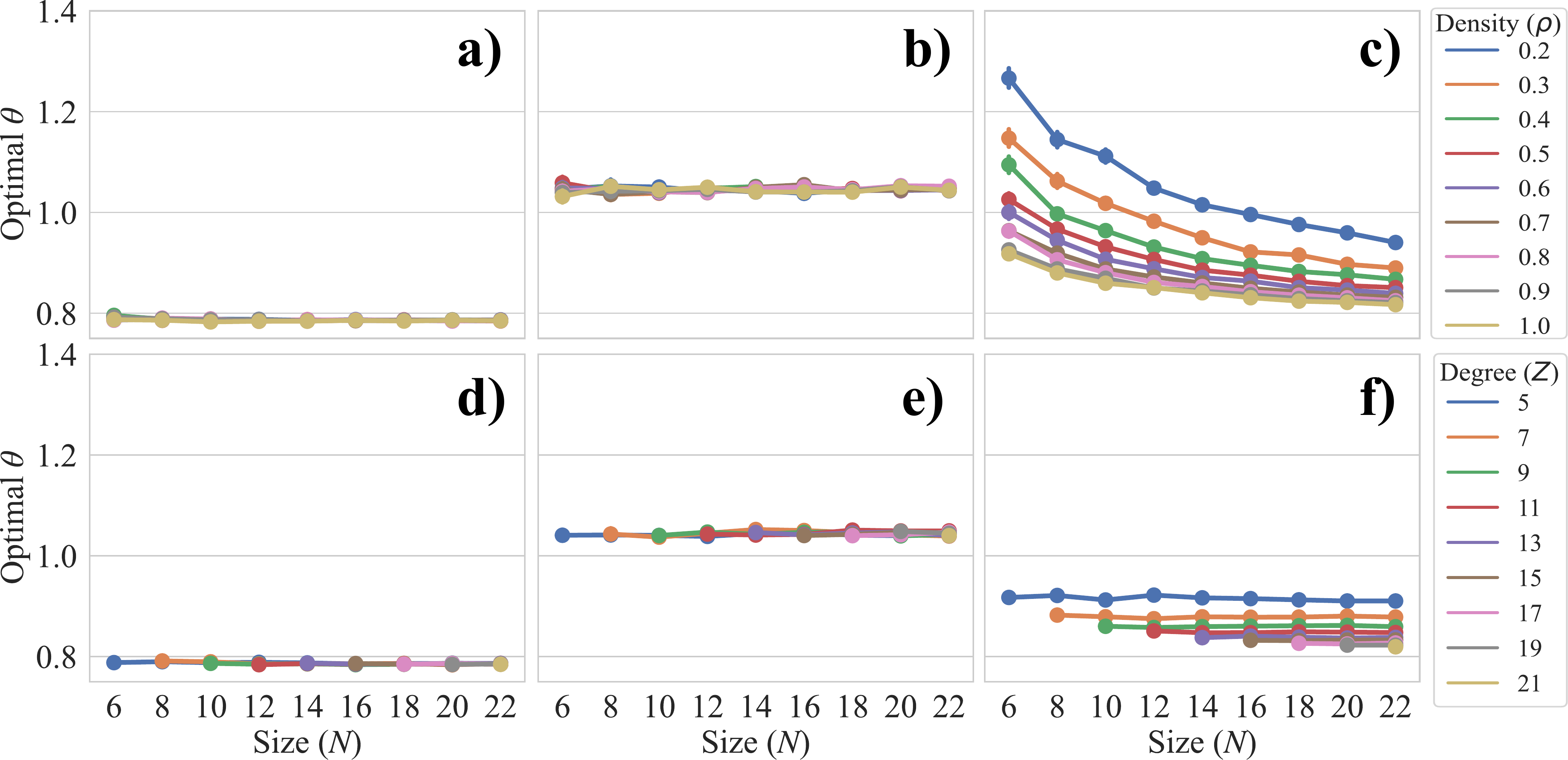}
\caption{Evidence that the optimal angle $\theta_{opt}$ of the single-layer QAOA converges to approximately a constant value. The results show the average of 500 instances with $\sigma^2=1$ and a $95\%$ confidence interval in the estimate of the average for (a,d) MaxCut ($\theta_{opt}\approx\frac{\pi}{4}$), (b,e) QUBO ($\theta_{opt}\approx\frac{\pi}{3}$), and (c,f) Random Ising model problems on (a,b,c) $G_{n,M}$ random graphs and (d,e,f) r-regular graphs. The trend is very clear for MaxCut and QUBO with any connectivity of the graph. In the case of Random Ising, the optimal $\theta$ seems to vary, mainly when the network is sparse (Color online).}
\label{Fig:thetafit}
\end{figure}

\begin{figure}[h]
\includegraphics[width=0.9\linewidth]{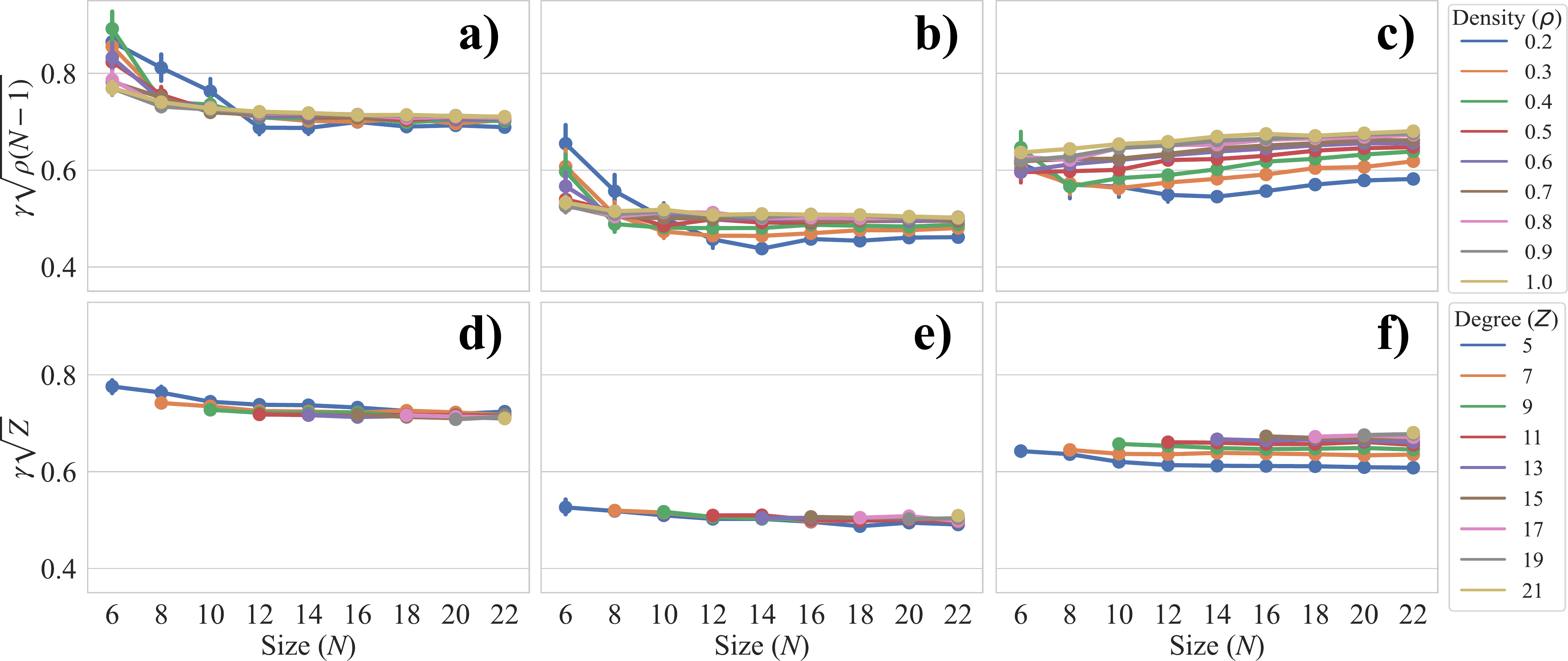}
\caption{Evidence that the optimal angle $\gamma_{opt}$ of the single-layer QAOA scales as $\approx \dfrac{\chi_{\gamma}}{\sigma\sqrt{(N-1)\cdot\rho}}=\dfrac{\chi_{\gamma}\sqrt{N}}{\sigma\sqrt{2M}}$. The results show the average of 500 instances with $\sigma^2=1$ and a $95\%$ confidence interval in the estimate of the average for (a,d) MaxCut, (b,e) QUBO, and (c,d) Random Ising model problems on (a,b,c) $G_{n,M}$ random graphs and (d,e,f) r-regular graph. The trend is especially noticeable with highly connected graphs (Color online).}
\label{Fig:gammaFit_randomgraph}
\end{figure}

\begin{figure}[h]
\includegraphics[width=0.9\linewidth]{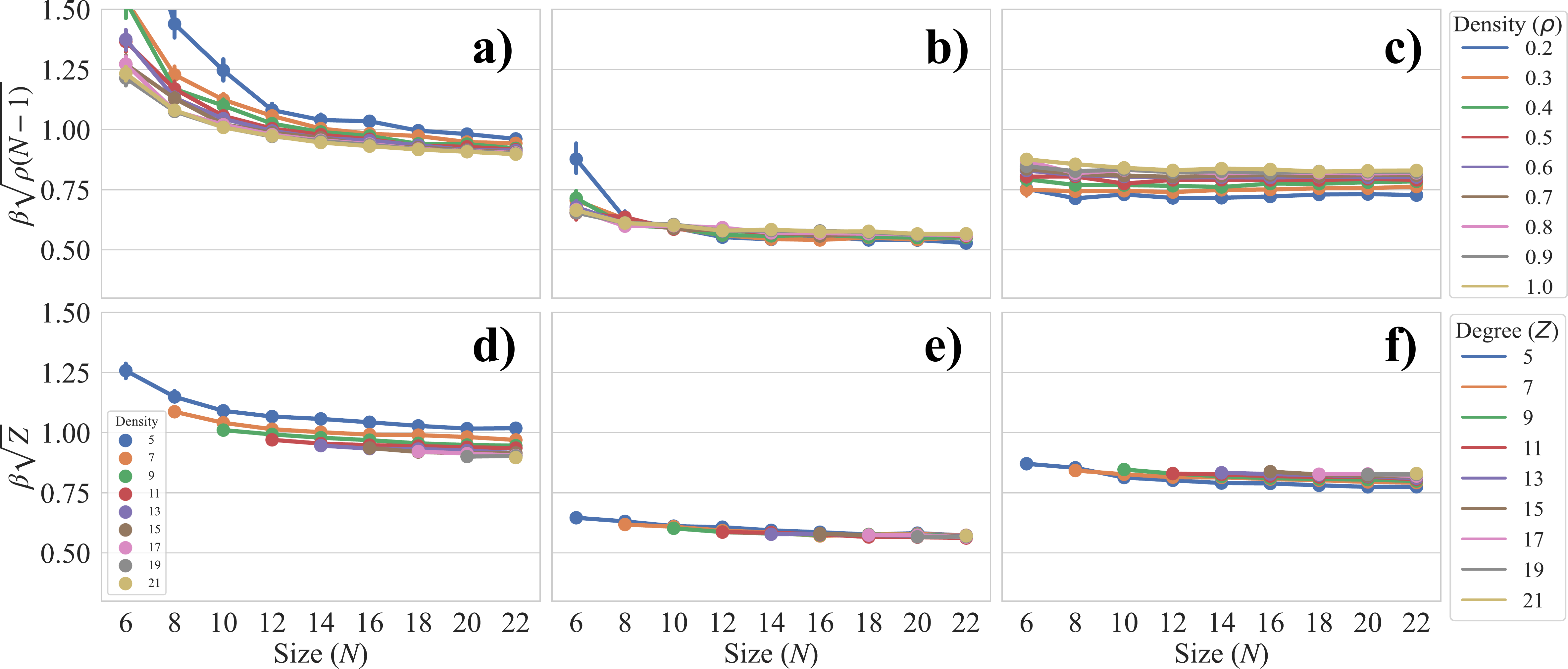}
\caption{Evidence that the average effective inverse temperature $\beta$ of the single-layer QAOA with optimal angles scales as $\approx \dfrac{\chi_{\beta}}{\sigma\sqrt{(N-1)\cdot\rho}}=\dfrac{\chi_{\beta}\sqrt{N}}{\sigma\sqrt{2M}}$ when $N\rightarrow\infty$. The results show the average of 500 instances with $\sigma^2=1$ and a $95\%$ confidence interval in the estimate of the average for (a,d) MaxCut, (b,e) QUBO, and (c,d) Random Ising model problems on (a,b,c) $G_{n,M}$ random graphs and (d,e,f) r-regular graph (Color online).}
\label{Fig:betaFit_randomgraph}
\end{figure}

\begin{figure}[h]
\includegraphics[width=0.9\linewidth]{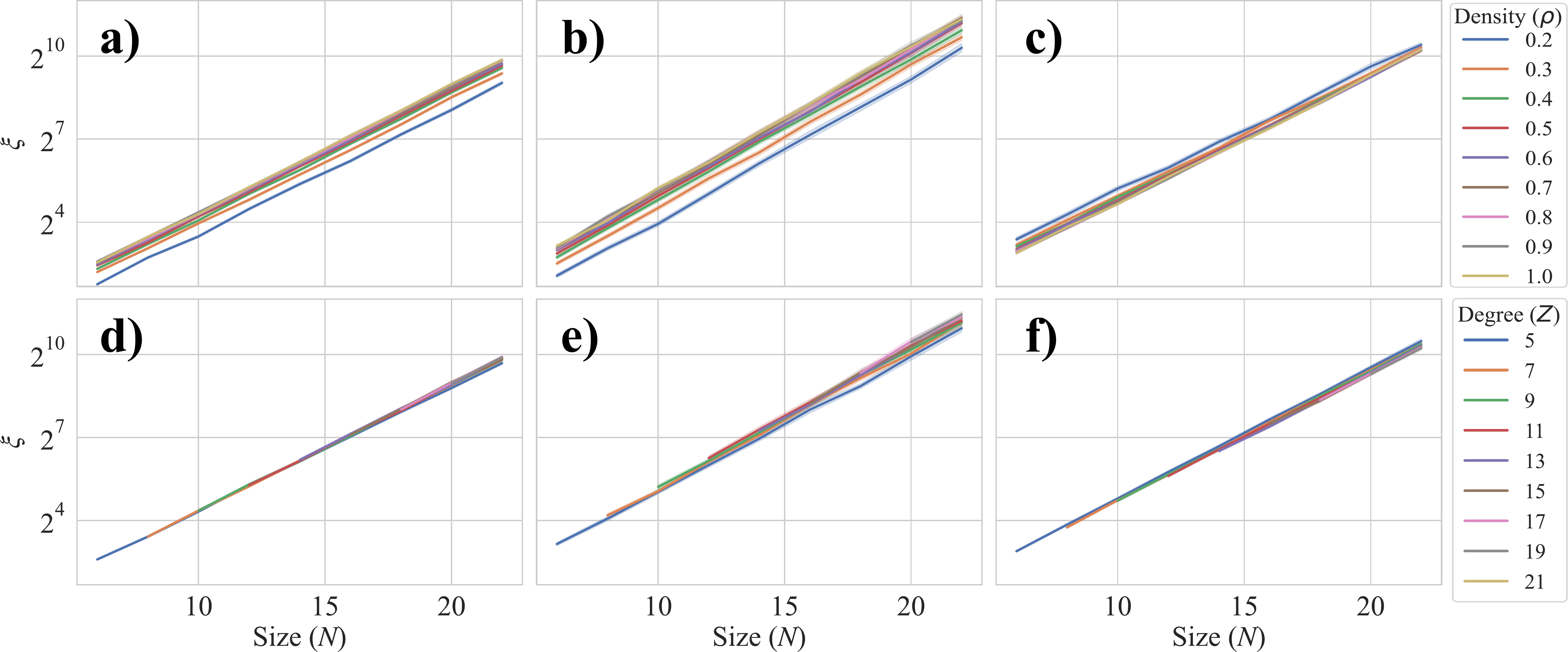}
\caption{Evidence that the enhancement in the ground state probability produced by the single-layer QAOA circuit with optimal angles $\xi$ scales as $\approx\sqrt{2^{N}}$. The results show the average of 500 instances with $\sigma^2=1$ and a $95\%$ confidence interval in the estimate of the average for (a,d) MaxCut, (b,e) QUBO, and (c,d) Random Ising model problems on (a,b,c) $G_{n,M}$ random graphs and (d,e,f) r-regular graph (Color online).}
\label{Fig:enhancementfit}
\end{figure}

We show in Figs. \ref{Fig:thetafit}-\ref{Fig:enhancementfit} the numerical evidence for the trends presented in table \ref{tab:scalings}. We include the results for all families of problems and random graph types analyzed in the paper. These plots do not show rigorous fits, and therefore some scalings may be slightly different. However, these results give a good idea of the pronounced tendency of these observables, namely optimal $\gamma$, optimal $\theta$, and the inverse temperature $\beta$ and the enhancement $\xi$ obtained with the optimal angles, where the optimal QAOA angles are those that minimize the average energy of the system.

\end{document}